\documentclass[aps,prc,twocolumn,amsmath,superscriptaddress,floatfix,nofootinbib]{revtex4-1}
\usepackage{setspace,ulem, epsfig,amssymb,amsfonts,amsmath,mathtools,bm,color,xcolor,graphicx,braket,adjustbox,esint,upgreek,verbatim}
\usepackage[bookmarksnumbered,bookmarksopen,colorlinks,citecolor=blue,linkcolor=red]{hyperref}
\usepackage{graphicx}  
\usepackage{booktabs}

\usepackage{float}
\usepackage{quantikz}
\usepackage{ctable}
\usepackage{subcaption} 
\usepackage{amsmath}
\allowdisplaybreaks

\graphicspath{
    {./figs-1/}  
    {./fig/} 
}

\begin{document}
\title{Quantum simulation of bottomonium dynamics in the quark–gluon plasma via the Lindblad equation}

\author{Linyuan Wei}
\email{linyuan\_wei@tju.edu.cn}
\affiliation{International Joint Institute of Tianjin University, Fuzhou,  Tianjin University, Tianjin  300072, China}

\author{Jin Hu}
 \email{hu-j23@fzu.edu.cn}
\affiliation{Department of Physics, Fuzhou University, Fujian 350116, China}

\author{Anping Huang}
 \email{huanganping425@cumt.edu.cn}
\affiliation{School of Material Science and Physics, China University of Mining and Technology, Xuzhou 221116, China}

\author{Yunpeng Liu}
 \email{yunpeng.liu@tju.edu.cn}
\affiliation{Department of Physics, Tianjin University, Tianjin 300354, China}

\author{Baoyi Chen}
 \email{baoyi.chen@tju.edu.cn}
\affiliation{Department of Physics, Tianjin University, Tianjin 300354, China}
\affiliation{International Joint Institute of Tianjin University, Fuzhou,  Tianjin University, Tianjin  300072, China}

\date{\today}
\begin{abstract}
Quantum computing provides a powerful framework for simulating real-time dynamics in open quantum systems, offering significant advantages for modeling heavy-quarkonium transport in high-energy nuclear collisions. In this work, we perform quantum simulations of the isotropic next-to-leading-order Lindblad equation for bottomonium in the quark-gluon plasma using a reduced spherical coordinate representation. We discretize operators and wavefunctions, map the physical state onto qubits, and execute time evolution via parameterized quantum gate operations. By extracting the $\Upsilon(1S)$ survival probability, we quantitatively isolate the color-octet contribution to the correlated regeneration of the bottomonium ground state and demonstrate that its overall impact on the final production of $\Upsilon(1S)$ in the hot QCD medium is small at temperatures accessible at the Large Hadron Collider. Furthermore, we have optimized the quantum simulation algorithm for the Lindblad equation. The improved algorithm requires only a single ancillary qubit to realize the Lindblad evolution, thereby significantly reducing the circuit complexity.

\end{abstract}

\maketitle

{\itshape Introduction ---} 
Ultrarelativistic heavy-ion collisions conducted at the Relativistic Heavy Ion Collider (RHIC) and the Large Hadron Collider (LHC) provide a unique experimental platform for investigating the quark-gluon plasma (QGP), a deconfined state of strongly interacting matter. The microscopic properties of the QGP are traditionally probed by a hierarchy of penetrating probes that traverse the hot and dense medium, among which heavy quarkonia serve as prime candidates~\cite{Apolinario:2022vzg,annurev:/content/journals/10.1146/annurev-nucl-121423-101041,Ghiglieri:2012iaw,Mocsy:2013syh,Ferreiro:2018umi,sym16020225,Rapp:2008tf}. Inside the QGP, heavy quarkonia undergo dissociation via color screening~\cite{MATSUI1986416,Mocsy:2007jz,Singh:2015eta} and inelastic scattering with thermal partons~\cite{PhysRevD.110.074040,Chen:2018dqg,Brambilla:2013dpa}. Concurrently, uncorrelated heavy-quark pairs diffusing through the medium can recombine into bound states via coalescence processes. To capture these complex dynamics, various theoretical frameworks—including transport models~\cite{Liu2022, Yao:2020eqy,Fu:2025opr}, coalescence approaches~\cite{Greco:2007nu}, statistical hadronization models~\cite{Andronic:2003zv,Andronic:2007bi,Andronic:2006ky,Braun-Munzinger:2009dzl,Andronic:2010dt,Andronic:2019wva}, and complex potential models~\cite{Wen:2022yjx,Dong:2022mbo}—have been developed to evaluate observables such as the nuclear modification factor $R_{AA}$ and elliptic flow coefficients $v_n$. Recently, open quantum system (OQS) frameworks based on the Lindblad equation~\cite{lindblad_generators_1976,gorini_completely_1976,Akamatsu:2014qsa,Armesto:2026fit} and quantum master equations~\cite{Barata:2023uoi,Borghini:2011ms,Miura:2019ssi,Delorme:2024rdo,BR:2025lhx} have been applied to the study of heavy quarkonium. These approaches naturally account for quantum coherence, thermal dissipation, and color exchange.

In high-energy nuclear physics, quantum algorithms have recently been applied to model jet quenching~\cite{Qian:2025fnx,Barata:2021hyh,Castro:2025ocx}, color decoherence~\cite{Barata:2026icn,Twagirayezu:2025der}, and heavy-quark diffusion~\cite{Du:2023ewh}. However, standard Hamiltonian simulation techniques inherently assume closed quantum systems governed by unitary dynamics, so simulating non-unitary Lindblad dynamics—which describes mixed states and non-norm-preserving operations—falls outside the scope of conventional Hamiltonian algorithms. To circumvent this limitation, dilation techniques can be employed to embed the non-unitary system evolution into a larger unitary operator acting on an extended Hilbert space supplemented by ancillary qubits~\cite{Cleve:2016dgx}.

In this work, we present a noise-free quantum simulation of the isotropic next-to-leading-order Lindblad equation for bottomonium states~\cite{Brambilla:2022ynh} using the Qiskit simulator. By explicitly incorporating both color-singlet and color-octet $b\bar{b}$ configurations, our simulation quantitatively isolates the impact of color-octet states on the final $\Upsilon(1S)$ survival probability and studies their contribution to the final production of $\Upsilon(1S)$ in the hot deconfined medium. Throughout this work, we adopt natural units with \(k_B = \hbar = c = 1\).

{\itshape Lindblad Equation for Heavy Quarkonium ---} 
By treating heavy quarkonium as an open quantum system in a hot QCD medium, its dynamical evolution can be described by the Lindblad equation~\cite{Islam:2025dvm,Brambilla:2022ynh,Akamatsu:2021dot,Akamatsu:2020ypb,Miura:2022arv,Strickland:2021cox,Brambilla:2020qwo,Brambilla:2016wgg,Brambilla:2017zei}.
\begin{align}
     \frac{d\rho}{dt} = -i[\bar{H},\rho] + \sum_{i=1}^{m}  \left( L_i \rho L_i^\dagger - \frac{1}{2} \{L_i^\dagger L_i, \rho\} \right),\label{Lindblad in reduce wavefunction}
\end{align}
where $\rho$ and $\bar{H}$ denote the density matrix and the Hamiltonian of the heavy quarkonium system, respectively. The Hamiltonian $\bar{H}$ contains color-singlet and color-octet components.
\begin{align}
    \bar{H} = \begin{pmatrix}
\bar{h}_s & 0 \\
0 & \bar{h}_o
\end{pmatrix},\label{block-diagonal Hamiltonian matrix}
\end{align}
which are defined in Ref.~\cite{Islam:2025dvm,Brambilla:2022ynh} as
\begin{align}
    \bar{h}_s &= \frac{\overline{\mathcal{D}}^2}{M} - \frac{C_f \alpha_s}{r} + \frac{\hat{\gamma} T^3}{2} r^2 + \frac{\hat{\kappa} T^2}{4M} \{r, p_r\},
\\
    \bar{h}_o &= \frac{\overline{\mathcal{D}}^2}{M} + 
\frac{1}{2N_c} \frac{\alpha_s}{r} + 
\frac{N_c^2 - 2}{2(N_c^2 - 1)}
\left[
\frac{\hat{\gamma} T^3}{2} r^2 + 
\frac{\hat{\kappa} T^2}{4M} \{r, p_r\}
\right],
\end{align}
with $\overline{\mathcal{D}}^2 = -\frac{\partial^2}{\partial r^2} + \frac{l(l+1)}{r^2}$. Here, $M = m_{\Upsilon (1S)}/2 = 4.73\ \text{GeV}$ is the reduced mass, $C_f = 4/3$ is the color factor, $\alpha_s = 0.468$ is the strong coupling constant, $N_c = 3$ is the number of colors, and $r$ denotes the relative distance between the heavy quark and antiquark. Additionally, $l$ is the orbital angular momentum quantum number, $p_r$ is the radial momentum operator, and $T$ is the temperature of the medium. The dimensionless coefficients $\hat{\kappa} = \kappa/T^3$ and $\hat{\gamma} = \gamma/T^3$ are defined, following Ref.~\cite{Brambilla:2022ynh}, in terms of the momentum diffusion coefficient $\kappa$ and the color-electric screening coefficient $\gamma$, respectively. The Lindblad operators $L_i = \gamma_i \overline{C}_i$, adopted from Ref.~\cite{Brambilla:2022ynh}, consist of the jump operators $\overline{C}_i$ and their corresponding dissipation rates $\gamma_i$,
\begin{align}
   \overline{C}^{\uparrow}_{s\to o} &= r - \frac{N_c \alpha_s}{8T} + \frac{1}{2MT} \left( \frac{\partial}{\partial r} - \frac{l+1}{r} \right), 
\\
\overline{C}^{\downarrow}_{s\to o} &= r - \frac{N_c \alpha_s}{8T} + \frac{1}{2MT} \left( \frac{\partial}{\partial r} + \frac{l}{r} \right), 
\\
\overline{C}^{\uparrow}_{o\to s} &= r + \frac{N_c \alpha_s}{8T} + \frac{1}{2MT} \left( \frac{\partial}{\partial r} - \frac{l+1}{r} \right), \\
\overline{C}^{\downarrow}_{o\to s} &= r + \frac{N_c \alpha_s}{8T} + \frac{1}{2MT} \left( \frac{\partial}{\partial r} + \frac{l}{r} \right), \\
\overline{C}^{\uparrow}_{o\to o} &= r + \frac{1}{2MT} \left( \frac{\partial}{\partial r} - \frac{l+1}{r} \right), \\
\overline{C}^{\downarrow}_{o\to o} &= r + \frac{1}{2MT} \left( \frac{\partial}{\partial r} + \frac{l}{r} \right),
\end{align}
and the corresponding dissipation rates $\gamma_{i}$~\cite{Brambilla:2022ynh} are  
\begin{align}
    \gamma^{\uparrow}_{s\to o} &= \sqrt{\hat{\kappa} T^3 \frac{l+1}{2l+1}}, \label{dissipation rates1}\\
    \gamma^{\downarrow}_{s\to o} &= \sqrt{\hat{\kappa} T^3 \frac{l}{2l+1}}, \\
    \gamma^{\uparrow}_{o\to s} & = \sqrt{ \frac{\hat{\kappa} T^3}{N_c^2 - 1} \frac{l+1}{2l+1}}, \\
     \gamma^{\downarrow}_{o\to s} &= \sqrt{ \frac{\hat{\kappa} T^3}{N_c^2 - 1} \frac{l}{2l+1}}, \\
     \gamma^{\uparrow}_{o\to o} &= \sqrt{\hat{\kappa} T^3 \frac{N_c^2 - 4}{2(N_c^2 - 1)} \frac{l+1}{2l+1}} ,\\
     \gamma^{\downarrow}_{o\to o} &= \sqrt{\hat{\kappa} T^3 \frac{N_c^2 - 4}{2(N_c^2 - 1)} \frac{l}{2l+1}}.  \label{dissipation rates}
\end{align}
The six Lindblad operators correspond to distinct internal transition channels within the quarkonium system: transitions from the color-singlet state to the color-octet state, from the color-octet state to the color-singlet state, and transitions among the color-octet states. Each of these transitions can be further classified according to whether it raises or lowers the orbital angular momentum quantum number \(l\) (\(\Delta l = \pm 1\)).
\begin{figure}[htbp]    
    \centering             
\includegraphics[width=0.35\textwidth]{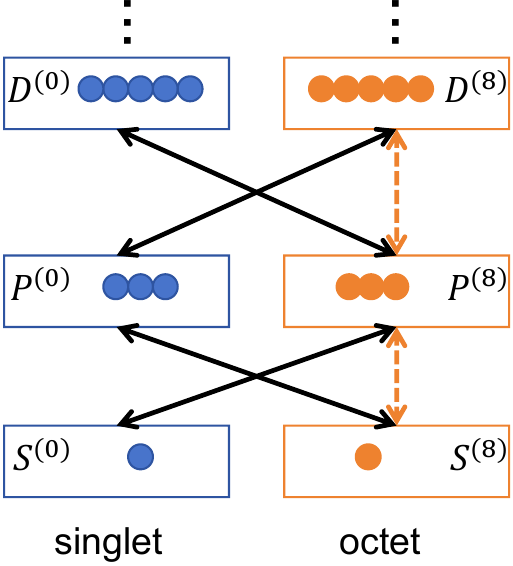}  
    \caption{Transitions between color-singlet and color-octet states. The solid black double-headed arrows indicate transitions between color-singlet and color-octet states, while the dashed orange double-headed arrows denote transitions within the color-octet states. The number of filled circles represents the degeneracy of each energy level.}  
    \label{color transitions}    
\end{figure}

{\itshape Quantum Circuit for Lindblad Equation ---} 
Quantum simulations have been successfully employed to solve the Lindblad equation~\cite{Cleve:2016dgx,DeJong:2020riy,PhysRevD.106.054508,PhysRevLett.125.010501,PhysRevResearch.7.023076}. 
The effective Hamiltonian \(J\) can be constructed using the Lindblad operators. The total wave function is then evolved step by step using the time-evolution operator \(\exp(-iJ\sqrt{\delta t})\), where \(\delta t\) denotes the time step. Following Ref.~\cite{Cleve:2016dgx}, we construct the effective Hamiltonian \(J\) from the Lindblad operators as,
\begin{align}
     J = \begin{pmatrix}
    0 & L_1^\dagger & \cdots & L_m^\dagger \\
    L_1 & 0 & \cdots & 0 \\
    \vdots & \vdots & \ddots & \vdots \\
    L_m & 0 & \cdots & 0
    \end{pmatrix} .\label{J matrix}
\end{align}
The total wave function $\ket{\psi_{\text{total}}}$ consists of medium and heavy quarkonium components, which are stored in separate registers, each comprising multiple qubits.
\begin{align}
    \ket{\psi_{\text{total}}} &= \ket{\text{medium}} \otimes \ket{\text{color}} \otimes \ket{l} \otimes \ket{r} \nonumber \\
    &= \ket{q_{n - 1} \dots q_2 q_1 q_0},
\end{align}
where $\ket{r}$ and $\ket{l}$ represent the spatial and angular momentum components of the wave function, respectively. The state $\ket{\text{color}}$ denotes the color degree of freedom, and $\ket{\text{medium}}$ describes the state of the medium. The total number of qubits in the different registers is $n$.

To initialize the quantum circuit, the heavy quarkonium is prepared in the $\Upsilon(1S)$ state,
\begin{align}
    \ket{\psi_0} =  \ket{\text{singlet}} \otimes \ket{l=0} \otimes \ket{u_{1S}},
\end{align}
where the radial part of the wave function is defined as
\begin{align}
    u_{1S}(r) = r R_{n=1,l=0}(r),
\end{align}
where $R_{nl}$ is the radial wave function of the $\Upsilon(1S)$. With this information, the initial density matrix of the total system at the initial time $t_0$ is given by 
\begin{align}
    \rho_{\text{tot}}(t_0) = \ket{0}\bra{0} \otimes \ket{\psi_0} \bra{\psi_0}.  \label{initual rho}
\end{align}

To simulate the dissipative evolution of heavy quarkonium governed by the Lindblad master equation, we implement a digital quantum simulation scheme leveraging system and ancillary qubit registers.
The quarkonium state is encoded into $n_{\rm quarkonium}$ qubits. To incorporate environment-induced dissipation, $n_{\text{aux}} = \lceil \log_2(m + 1) \rceil$ ancillary qubits are introduced, where $m$ denotes the number of Lindblad jump operators.

At each discrete time step $\delta t$ with a local temperature $T$, the system-environment dynamics are implemented via two sequential quantum gate operations.
First, the dissipation gate $e^{-iJ\sqrt{\delta t}}$, constructed from the effective Hamiltonian $J$, is applied across the combined system and ancillary registers to induce quarkonium-environment entanglement.
Second, the unitary Hamiltonian gate $e^{-i\bar{H}\delta t}$ is applied exclusively to the quarkonium register.
The complete quantum circuit is transpiled into elementary gate sets for execution.
Following the unitary evolution, the environment is integrated out by taking the partial trace over the ancillary qubits: $\rho_{\text{sys}}(t + \delta t) = \operatorname{Tr}_{\text{aux}}(\rho_{\text{tot}})$.
The ancillary qubits are then reset to $\ket{0}$ for the next iteration.
This cycle is repeated until the local temperature of the medium falls below a critical value $T_f$.
Upon completion of the time evolution, the survival probability of the $\Upsilon(1S)$ state at time $t$ is extracted via
\begin{equation}
P_{\text{surv}}(t) = \operatorname{Tr}\big[ \mathcal{P}_\psi \rho_{\text{sys}}(t) \big],
\end{equation}
where $\mathcal{P}_\psi = \vert{}\psi\rangle\langle\psi\vert{}$ denotes the projection operator onto the $\Upsilon(1S)$ state.
A schematic diagram of the quantum circuit described above is shown in Fig.~\ref{fig: quantum circuit}.

\begin{figure}[htbp]
    \centering
    \scalebox{0.8}{
    \begin{quantikz}
       \lstick{$\ket{\psi}$} &   \gate[2]{e^{-i J \sqrt{\delta t}}} & \gate{e^{-i \bar{H} \delta t}}        &  \gate[2]{...}& \gate[2]{e^{-i J \sqrt{\delta t}}} & \gate{e^{-i \bar{H} \delta t}}  & \meter{}\\
        \lstick{$\ket{0}$} &         \qw                            & \gate{\ket{0}}                   & \qw  &        \qw                        & \gate{\ket{0}}            & \qw 
    \end{quantikz}}
    \caption{The ancillary register is initialized to \(|0\rangle\), and the system register is prepared in the state \(|\psi\rangle\). Boxes represent quantum gates acting on the corresponding qubits. A box labeled \(|0\rangle\) denotes a gate that resets the associated qubits to the \(|0\rangle\) state. The evolution of the Lindblad equation over the interval \([0, t]\) is realized by sequentially applying the quantum circuit for each time step \(\delta t\). In the limit \(\delta t \to 0\), the circuit outcome approaches the exact evolution of the Lindblad equation. Finally, the evolution results are obtained by measuring the states of the qubits in the system register.}
    \label{fig: quantum circuit}
\end{figure}
{\itshape Quantum Circuit Model Testing ---}
We use the well-established classical Lindblad evolution simulation package QuTiP as a benchmark to verify the accuracy of the numerical results obtained from the quantum circuit model constructed in this work. In this benchmark test, we use \(10\) qubits to encode the heavy quarkonium system: \(7\) qubits encode the relative coordinate \(r\), covering \(128\) discrete internal points; \(2\) qubits encode the orbital angular momentum, truncated at \(l_{\mathrm{max}} = 3\); and \(1\) qubit encodes the color state. With three additional ancillary qubits, the total number of qubits required is \(13\). Although \(u_{1\mathrm{S}}(r)\) is already sufficiently small at \(r = 5a_0\), where \(a_0 = 2/(C_f \alpha_s M)\)~\cite{Brambilla:2022ynh} is the Bohr radius of the bottomonium system, we set \(L = 40a_0\) to effectively suppress errors from boundary reflections of the evolved wavefunction. Details of the discretization and quantum information encoding are provided in Appendix~\ref{Discretization and Quantum Information Encoding}. To investigate the quarkonium dynamics in a cooling medium, the medium temperature is assumed to follow Bjorken hydrodynamics according to the relation
\begin{align}
T(t) = T_0 \left( \frac{t_0}{t} \right)^{1/3},\label{Bjorken}
\end{align}
where $t_0 = 0.6~\text{fm}/c$~\cite{CHATTERJEE2009503c} denotes the initial time of the Lindblad evolution, and $T_0 = 0.51~\text{GeV}$ represents the initial medium temperature~\cite{Csanad:2011jq}, estimated from hydrodynamic simulations of the hot, deconfined medium produced in nucleus-nucleus collisions at LHC energies.

\begin{figure}[htbp]    
    \centering             
\includegraphics[width=0.48\textwidth]{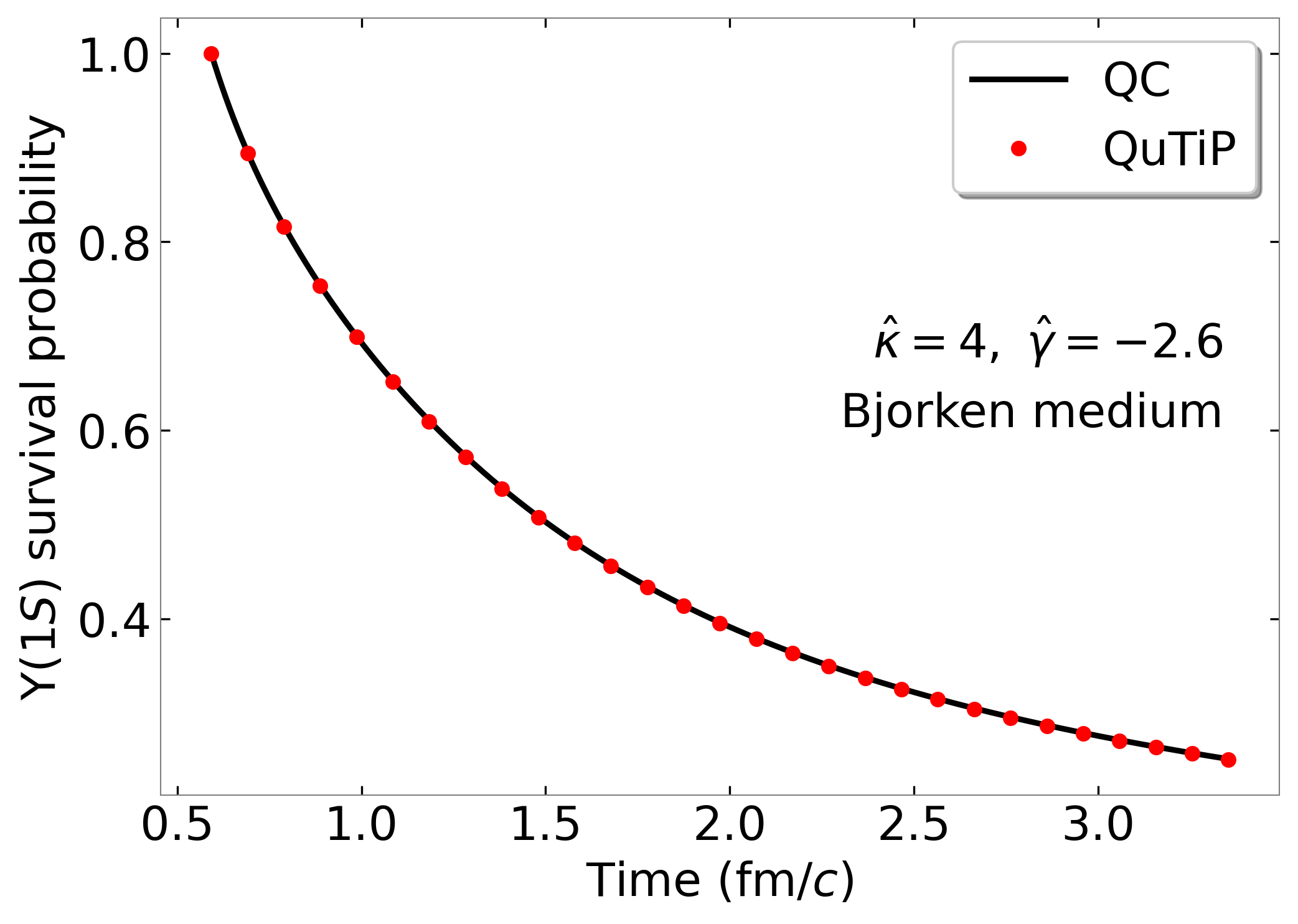}  
    \caption{Comparison of the quantum computing simulation results (line) and QuTiP results (dots)~\cite{qutip5,johansson2013qutip,johansson2012qutip}. The parameters in the Lindblad operators are taken as $\hat{\kappa}=4$ and $\hat{\gamma}=-2.6$ in both simulations.  }  
    \label{qc_qutip}    
\end{figure}

Figure~\ref{qc_qutip} presents the survival probabilities computed using both the traditional QuTiP method and the quantum circuit (QC) approach. The QC results are in excellent agreement with those from the classical QuTiP solver.

{\itshape Resource-Optimized Algorithm and Numerical Simulation Results ---}
The limited number of qubits restricts the application of quantum algorithms for solving the Lindblad equation. Thus, minimizing qubit requirements is essential for simulating the evolution of heavy quarkonium. In Ref.~\cite{PhysRevResearch.6.043321}, Lindblad dynamics is simulated using only a single ancillary qubit, achieved by resetting the bath qubit after each time step. Following this idea, we optimize the existing quantum algorithm.
In this algorithm, the effective Hamiltonian $J$ can be decomposed into
\begin{align}
    J = \sum_{i=1}^{m} \big( |i\rangle\langle 0| \otimes L_i + |0\rangle\langle i| \otimes L_i^\dagger \big) \equiv \sum_{i=1}^{m} J_i,
\end{align}
where each \(J_i\) couples only the states \(|0\rangle\) and \(|i\rangle\) within the ancillary register, and yields zero when acting on any other ancillary state \(|j\rangle\) (\(j \neq i\)).
For a sufficiently small time step \(\delta t\), the exponential operator \(e^{-iJ\sqrt{\delta t}}\) can be approximated using the first-order Trotter decomposition as (for more details, see~\cite{Trotter1959,Suzuki1976})
\begin{align}
    e^{-iJ\sqrt{\delta t}}  \approx \prod_{i=1}^{m} e^{-iJ_i \sqrt{\delta t}}.
\end{align}
The Trotter decomposition is first employed because individual gates $e^{-iJ_i \sqrt{\delta t}}$ are easier to transpile into elementary gate operations than the full operator $e^{-iJ \sqrt{\delta t}}$. Notably, each $e^{-iJ_i \sqrt{\delta t}}$ gate interacts exclusively with a specific environment state $\ket{i}$. In the original algorithm, after sequentially applying these decomposed $e^{-iJ_i \sqrt{\delta t}}$ gates, all environment states $\ket{i}$ are reset simultaneously. As a result, encoding the environment requires $n_{\text{aux}} = \lceil \log_2(m + 1) \rceil$ ancillary qubits, where $m$ is the number of Lindblad operators. However, if the environment is reset immediately following the application of each $e^{-iJ_i \sqrt{\delta t}}$ gate, this Lindblad quantum simulation algorithm can be executed using only a single ancillary qubit. Here, $J_i$ is represented as a $2\times2$ block matrix in terms of $L_i$.
\begin{align}
  \bar{J}_i   = \begin{bmatrix}
        0 & L_i^\dagger\ \\ L_i & 0
    \end{bmatrix},
\end{align}
This modification reduces both the qubit resources and the gate complexity of the quantum circuit, as detailed in Appendix~\ref{Circuit Gate Complexity: Original vs. New Algorithm}.

The corresponding quantum circuit for the optimized algorithm is shown in Fig.~\ref{fig: J gate}.
\begin{figure}[htbp]
    \centering
    \begin{quantikz}
       \lstick{$\ket{\psi}$} &  \gate[2]{e^{-i \bar{J}_1 \sqrt{\delta t}}} & \qw                                & \gate[2]{e^{-i  \bar{J}_2 \sqrt{\delta t}}}       & \qw                                & \gate[2]{...}\\
        \lstick{$\ket{0}$}         &    \qw                         & \gate[style={fill=white}]{\ket{0}} & \qw                                  & \gate[style={fill=white}]{\ket{0}}&        
    \end{quantikz}
    \caption{Schematic diagram of the quantum circuit for implementing \(e^{-iJ\sqrt{\delta t}}\) using a single ancillary qubit. Unlike Fig.~\ref{fig: quantum circuit}, \(\ket{0}\) here denotes only an ancillary qubit.}
    \label{fig: J gate}
\end{figure}
The results for the evolution of the \(\Upsilon(1S)\) survival probability using the optimized quantum algorithm under the same parameters are shown in Fig.~\ref{qc_single_compare}. These results agree with those obtained from the original algorithm within the error margin. 
\begin{figure}[htbp]    
    \centering             
    \includegraphics[width=0.48\textwidth]{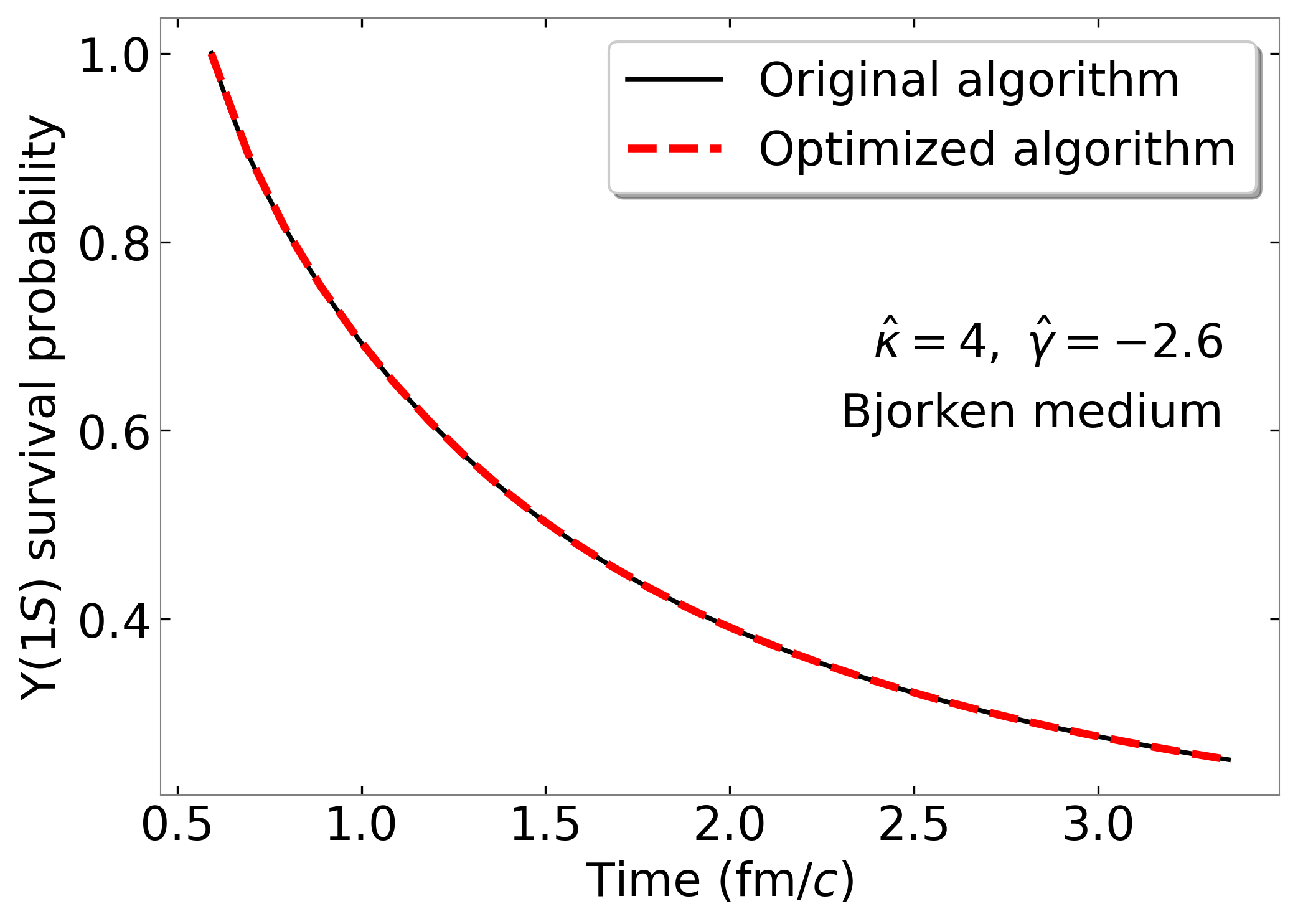}  
    \caption{Comparison of the $\Upsilon(1S)$ evolution results between the original and optimized algorithms in quantum circuits.}  
    \label{qc_single_compare}    
\end{figure}
Since the new algorithm reduces both the number of qubits and the gate complexity of the overall circuit, it lowers the computational cost of the Qiskit simulation and thereby enables longer physical evolution times for the heavy-quarkonium system. To determine an appropriate time step, we first performed a time-step convergence test using QuTiP, with the results shown in Fig.~\ref{delta_t error}. The results indicate that \(\delta t = 0.0197\ \mathrm{fm}/c\) provides sufficient convergence for the present long-time simulation. For the same physical evolution time, the smaller step size \(\delta t = 0.00197\ \mathrm{fm}/c\) requires ten times as many time steps. Taking all factors into account, we adopt a time step of \(\delta t = 0.0197\ \mathrm{fm}/c\) for this long-time simulation. The simulation runs until a hadron freeze-out temperature of approximately \(T_f \simeq 0.17\ \mathrm{GeV}\) is reached, with all other parameters identical to those used in the benchmark test.
\begin{figure}[htbp]    
    \centering             
    \includegraphics[width=0.48\textwidth]{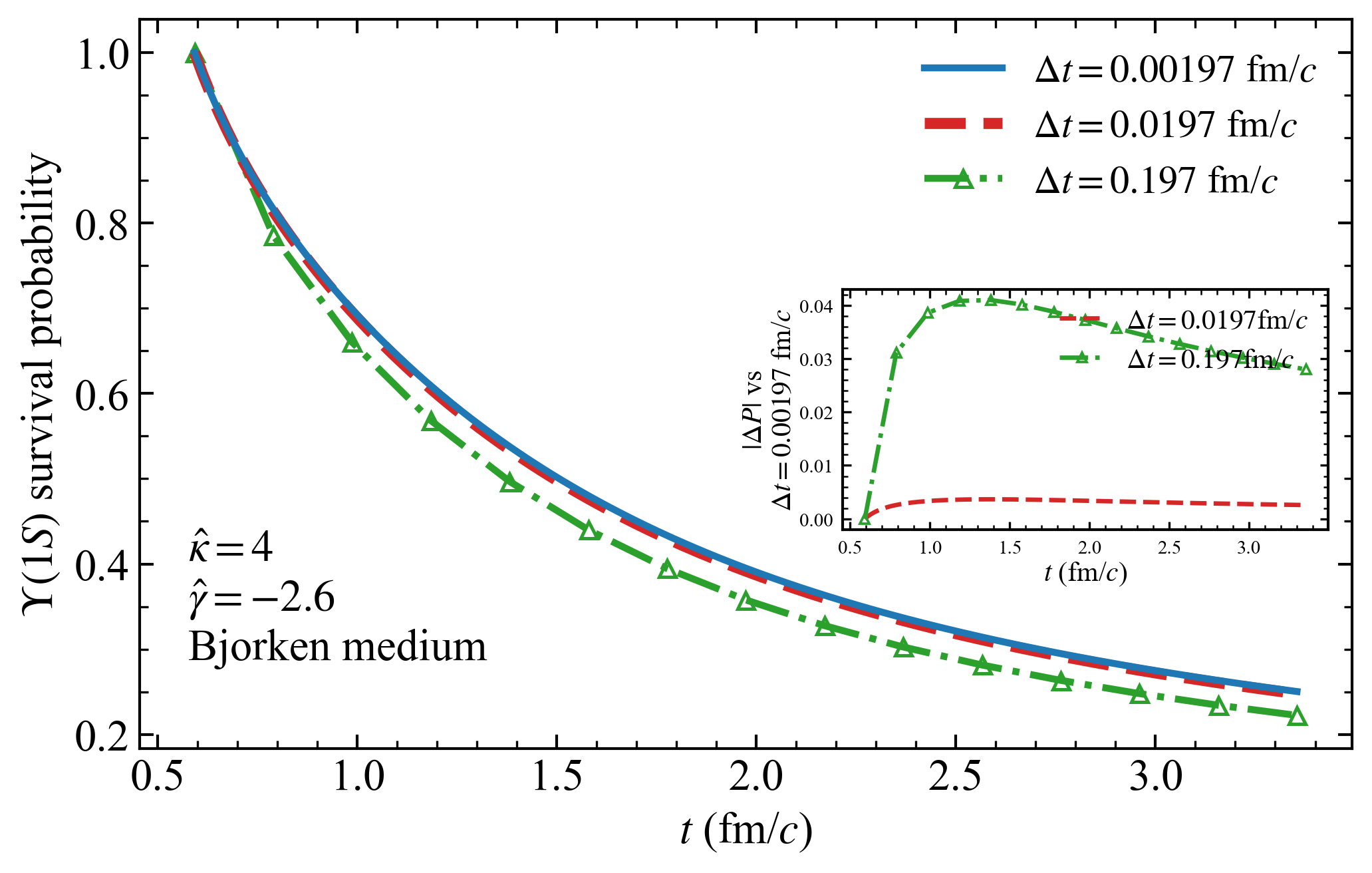}  
    \caption{Time-step convergence test using QuTiP.}  
    \label{delta_t error}    
\end{figure}
Leveraging our quantum circuit implementation, we investigate the contribution of color-octet to color-singlet transitions to the final $\Upsilon(1S)$ yield---a mechanism essential for understanding heavy quarkonium dynamics in heavy-ion collisions. Figure~\ref{s0andclose} compares the survival probability of $\Upsilon(1S)$ in a Bjorken-expanding medium under two scenarios: with and without octet-to-singlet transitions. Since the color-octet P-wave state ($l = 1$) is the only direct regeneration source for the color-singlet S-wave state ($l = 0$) in our model, we disable only this transition in the simulation, while keeping all other octet-to-singlet channels active. The close alignment between the two curves indicates that this transition channel has a negligible impact on the final $\Upsilon(1S)$ survival probability, which is in general agreement with the results shown in Fig.~3 of Ref.~\cite{Brambilla:2022ynh}.
\begin{figure}[htbp]    
    \centering             
    \includegraphics[width=0.48\textwidth]{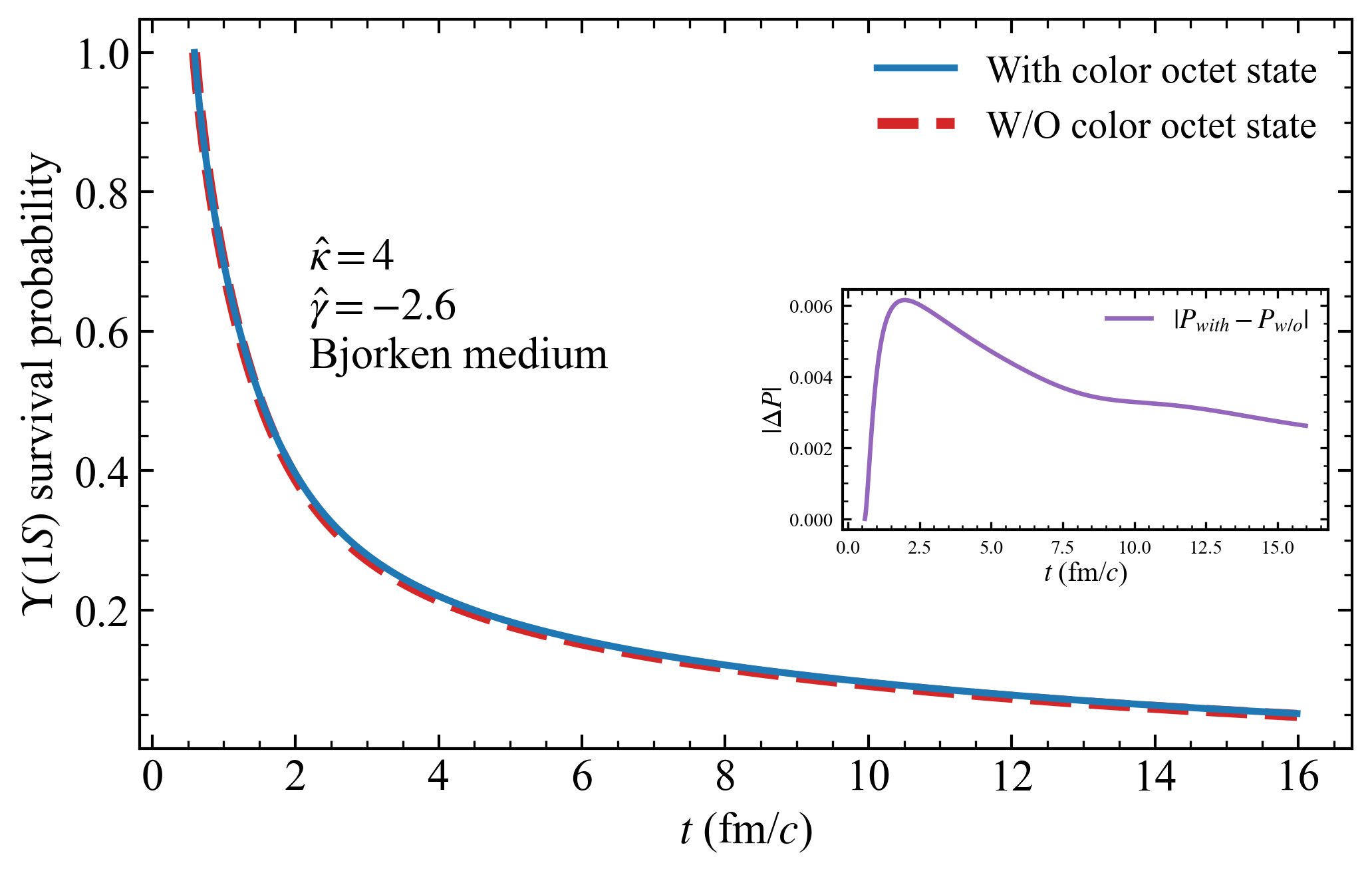}  
    \caption{The $\Upsilon(1S)$ survival probability in the Bjorken medium, with and without transitions from color-octet to color-singlet states, is plotted as solid and dashed lines, respectively. These calculations are based on the quantum circuit.}  
    \label{s0andclose}    
\end{figure}

To further investigate the contribution of color-octet states to \(\Upsilon(1S)\) regeneration, we perform two simulations, initializing the system in a color-octet S-wave state and a color-octet P-wave state, respectively. In each simulation, we examine the \(\Upsilon(1S)\) yield and the time evolution of the populations of the color-singlet and color-octet states.
\begin{figure}[htbp]    
    \centering             
    \includegraphics[width=0.48\textwidth]{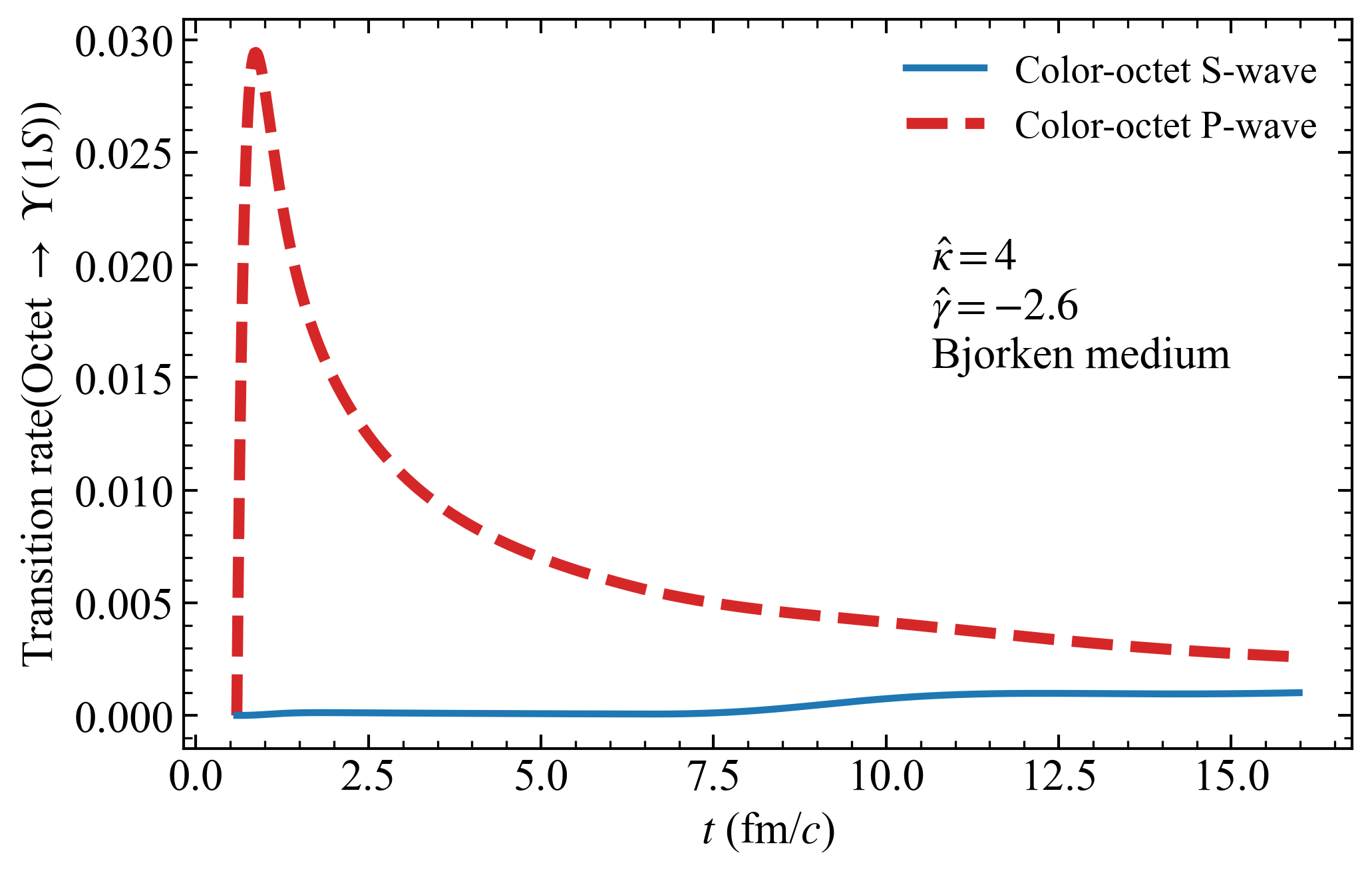}  
    \caption{We compute the transition probability from a color-octet state to a color-singlet $\Upsilon(1S)$ state in the Bjorken medium. These calculations are based on the quantum circuit.}  
    \label{o0ando1}    
\end{figure}
The results in Fig.~\ref{o0ando1} show that in the early stage of evolution, the contribution of the color-octet P-wave to \(\Upsilon(1S)\) is much larger than that of the color-octet S-wave, as expected. This is because, in our model, the color-octet P-wave is the only direct regeneration source for the color-singlet S-wave; the color-octet S-wave must first transition to the color-octet P-wave before it can contribute to the regeneration of the color-singlet S-wave. Moreover, the color-octet P-wave has four available transition channels. By comparing the dissipation rates in Eqs.~\eqref{dissipation rates1}--\eqref{dissipation rates} ($\gamma^{\downarrow}_{o\to s} < \gamma^{\uparrow}_{o\to s}<\gamma^{\downarrow}_{o\to o} < \gamma^{\uparrow}_{o\to o} <\gamma^{\downarrow}_{s\to o} < \gamma^{\uparrow}_{s\to o}$), one finds that transitions from color-singlet to color-octet states are much more probable than the reverse transitions, and that transitions preferentially occur toward higher energy levels. Among the four transition channels of the octet P-wave, the transition rate to the color-singlet S-wave is the smallest. Thus, internal transitions within the color-octet sector are more probable than transitions to the color-singlet. In addition, the repulsive potential of the color-octet state drives the \(b\bar{b}\) pair toward larger relative separation \(r\), which further disfavors the regeneration of \(\Upsilon(1S)\) \cite{Brambilla:2022ynh}. Consequently, in the intermediate stage of evolution, the \(\Upsilon(1S)\) initially produced through the octet P-wave exhibits an evolution curve similar to that in Fig.~\ref{s0andclose}, albeit with smaller values. Meanwhile, the contribution of the color-octet P-wave gradually approaches that of the color-octet S-wave. Eventually, the population distributions approach equilibrium, as shown in Figs.~\ref{o0} and~\ref{o1}.
\begin{figure}[htbp]    
    \centering             
    \includegraphics[width=0.48\textwidth]{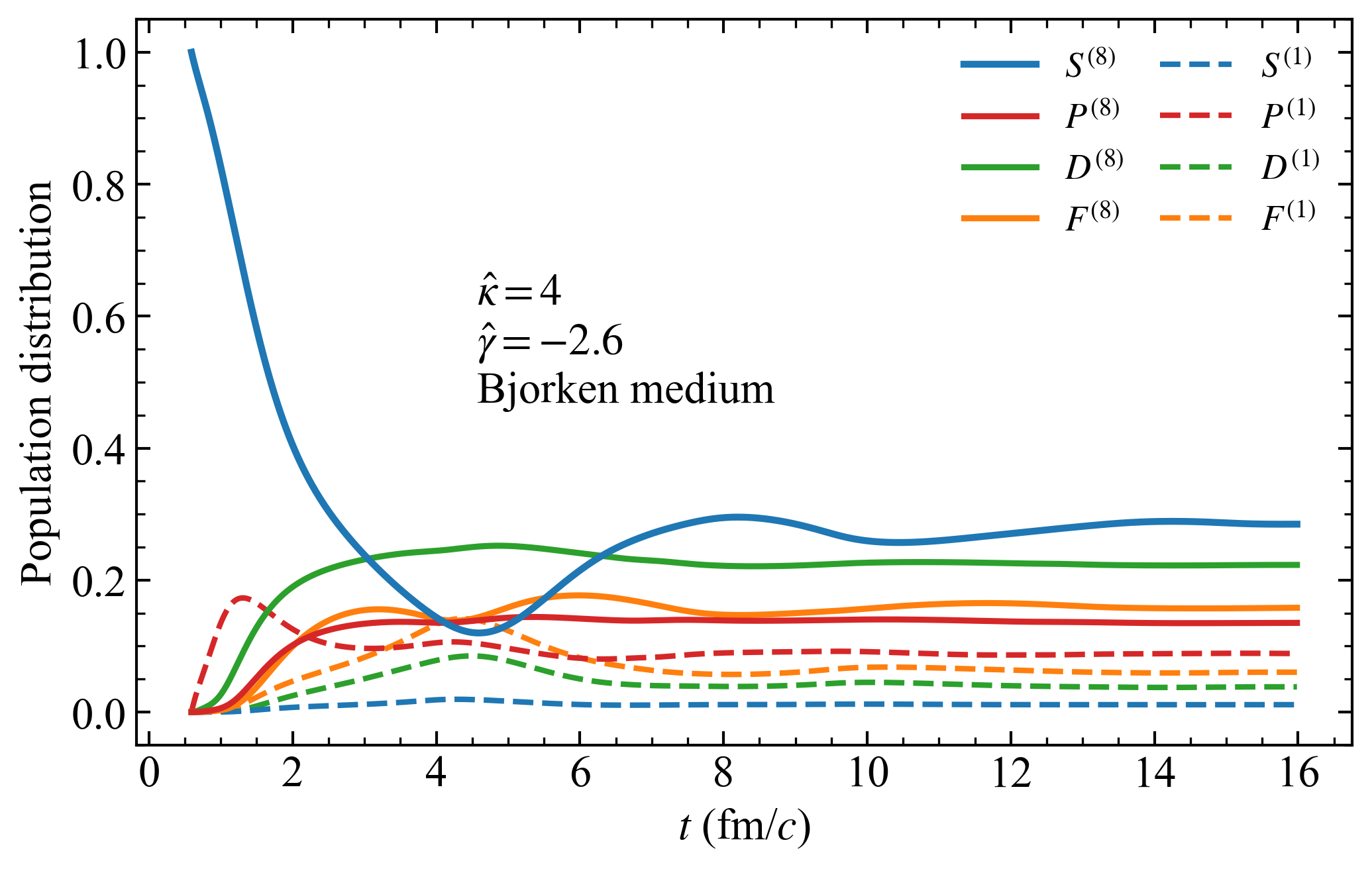}  
    \caption{Evolution of the population distribution for the initial color-octet S-wave state $S^{(8)}$}  
    \label{o0}    
\end{figure}
\begin{figure}[htbp]    
    \centering             
    \includegraphics[width=0.48\textwidth]{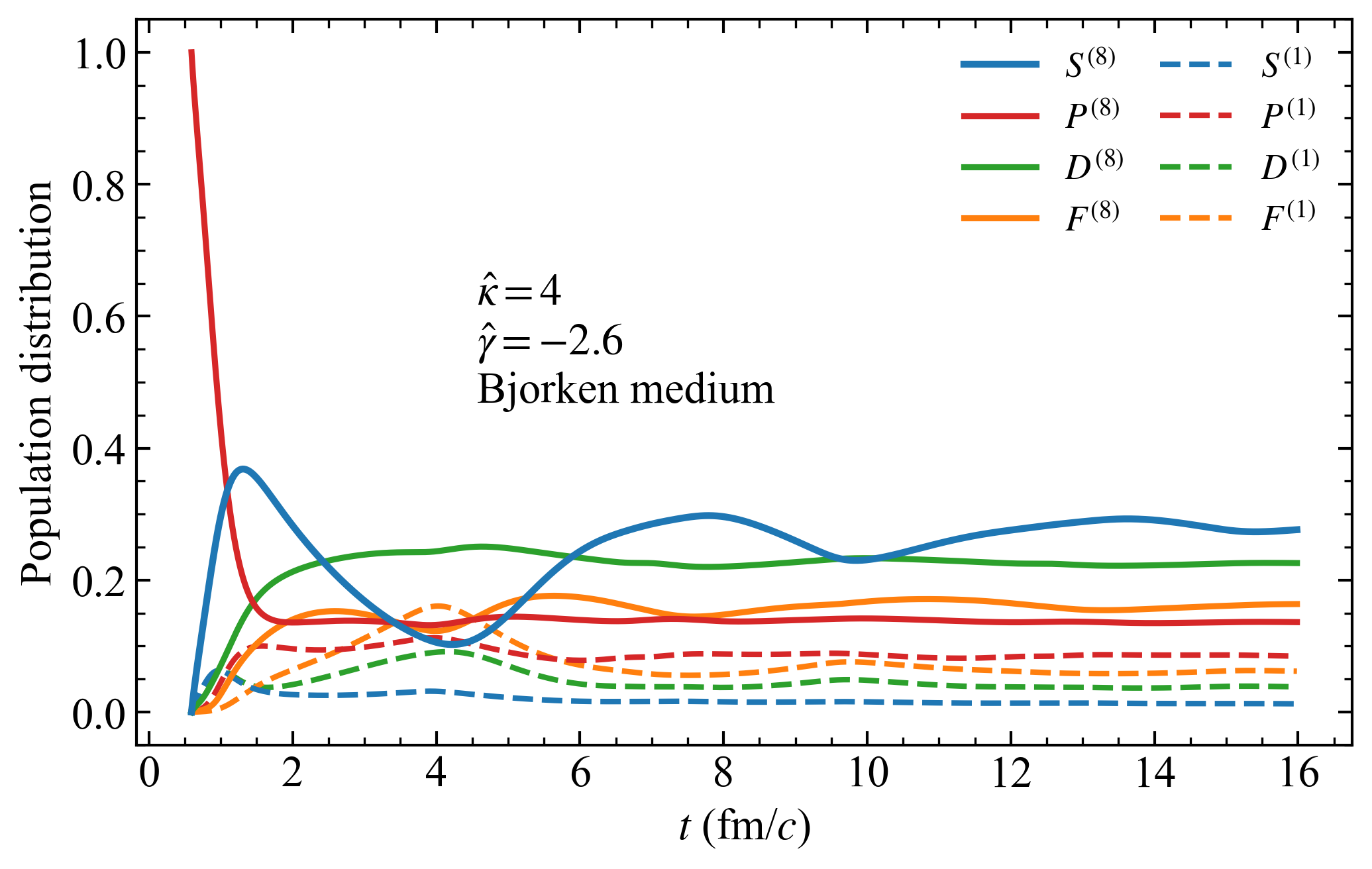}  
    \caption{Evolution of the population distribution for the initial color-octet P-wave state $P^{(8)}$.}  
    \label{o1}    
\end{figure}
As shown in Fig.~\ref{o0}, the time evolution of the population of the color-octet S-wave state \(S^{(8)}\) exhibits a pronounced non-monotonic behavior. During the early stage (\(t < 4.5\ \mathrm{fm}/c\)), the color-octet S-wave state is rapidly depleted through thermal excitation to higher orbital angular momentum states, thermal dissociation, and transitions to color-singlet states, causing a rapid decrease in its population. As the system cools due to Bjorken expansion, the thermal dissociation rate of the color-octet S-wave state decreases substantially. At \(t \approx 4.5\ \mathrm{fm}/c\) (corresponding to a temperature of about \(0.259\ \mathrm{GeV}\)), the population reaches a local minimum before undergoing a pronounced revival. This revival is predominantly attributed to feed-down from the previously populated higher orbital angular momentum states via de-excitation and color recombination during the cooling process. A similar evolution is also observed for the color-octet S-wave state in Fig.~\ref{o1}.

The theoretical model adopted in this paper and in related studies considers only the contribution of color-octet states to the correlated regeneration of \(\Upsilon(1S)\) during the evolution of the medium under Bjorken expansion. The results indicate that this regeneration contribution to the \(\Upsilon(1S)\) yield is relatively small. This result is consistent with Fig.~3 of Ref.~\cite{Brambilla:2022ynh}, but differs from Figs.~2 and 3 of Ref.~\cite{Zheng:2026alo}; the discrepancy arises from the different imaginary part of the potential and the isotropic assumption for \(\hat{C}^0\) adopted in that model. In a realistic QGP fireball, the contribution of uncorrelated regeneration should also be taken into account.

{\itshape Challenges for Implementation on Real Quantum Devices ---}
Although we have implemented the simulation of this Lindblad open quantum system on the Qiskit quantum simulation platform, many pressing challenges remain before Lindblad dynamics can be reliably simulated on real quantum devices. First, all simulations in this work are performed in a noise-free environment, whereas simulations on real quantum hardware inevitably suffer from noise, which typically must be addressed via quantum error mitigation techniques~\cite{RevModPhys.95.045005,Aharonov:2025wct} or quantum error correction codes~\cite{202509.2149,erol_comprehensive_2025}. Second, obtaining the full density matrix on real hardware requires quantum state tomography~\cite{PhysRevLett.74.4101,Zhao:2025lmd}. This process involves repeated state preparation, a set of measurements, and statistical inference, inevitably introducing statistical errors. More importantly, the gate depth of current quantum circuits far exceeds the capability of existing devices; therefore, more advanced transpilation schemes~\cite{Yan:2024upq,Decker:2025jts,Decker:2025juu} and more efficient quantum algorithms~\cite{PhysRevResearch.6.043321,PhysRevResearch.7.023076} are urgently needed. These challenges remain active research frontiers in the field.

{\itshape Conclusion ---}
In this work, we investigate the dynamical evolution of bottomonium in a hot, deconfined medium by modeling it as an open quantum system governed by the Lindblad equation. A quantum circuit framework is implemented to simulate the Lindblad dynamics. The results obtained from the quantum circuit show excellent agreement with those from the traditional QuTiP solver, validating the reliability and accuracy of our quantum approach. Furthermore, we leverage this framework to evaluate the contribution of color-octet states to the final yield of the bottomonium ground state \(\Upsilon(1S)\) through correlated regeneration. Our calculations reveal that octet-to-singlet transitions have a negligible effect on the correlated regeneration contribution to the final \(\Upsilon(1S)\) yield. Furthermore, an optimized algorithm is also tested for solving the Lindblad equation using one qubit to store the medium information. Overall, this work demonstrates that quantum computing provides a promising and viable pathway for simulating open quantum systems in high-energy nuclear physics.

\vspace{1cm}
{\bf Acknowledgement:} LW and BC are supported by the National Natural Science Foundation of China (NSFC) under Grant Nos. 12575149 and 12175165. JH is supported by the NSFC under Grant No.~12505149.

\bibliographystyle{apsrev4-1}
\bibliography{ref}

\appendix
\numberwithin{equation}{section}

\section{Discretization and Quantum Information Encoding}\label{Discretization and Quantum Information Encoding}
We discretize the interval \([0, L]\) into \(N+1\) equally spaced points \(r_j = j \Delta r\), where \(\Delta r = L/N\) and \(j = 0, 1, \dots, N\). For the wavefunction evolution, we impose Dirichlet boundary conditions, under which the distribution function vanishes at the boundary points and carries no effective information. Hence, only the interior points need to be represented. The interior points are \(j = 1, 2, \dots, N-1\), giving a total of \(\bar{N} = N-1\) points. The discretization matrices below are defined for the interior points, i.e., the matrix dimensions are \(\bar{N} \times \bar{N}\).

\textbf{1. Operator \(r\):} A diagonal matrix whose diagonal entries are the values of \(r\) at the interior points:
\begin{align}
     R_{jj} = r_j = j \Delta r, \quad j = 1, \dots, \bar{N}.
\end{align}
   All off-diagonal entries are zero.

\textbf{2. Operator \(1/r\):} A diagonal matrix whose diagonal entries are the values of \(1/r\) at the interior points:
\begin{align}
     R^{-1}_{jj} = \frac{1}{r_j} = \frac{1}{j \Delta r}, \quad j = 1, \dots, \bar{N}.
\end{align}
   Note that the point \(r=0\) is singular, but since it is not an interior point, it does not affect the matrix.

\textbf{3. Operator \(\partial_r\) (first derivative):} A central difference scheme is adopted. The matrix \(D_1\) is defined as:
   \begin{align}
         (D_1 u)_j \approx \frac{u_{j+1} - u_{j-1}}{2 \Delta r}, \quad j = 1, \dots, \bar{N},
   \end{align}
   i.e.,
\begin{align}
      D_1 = \frac{1}{2\Delta r}
   \begin{bmatrix}
   0 & 1 & 0 & \cdots & 0 \\
   -1 & 0 & 1 & \cdots & 0 \\
   0 & -1 & 0 & \ddots & \vdots \\
   \vdots & \ddots & \ddots & \ddots & 1 \\
   0 & \cdots & 0 & -1 & 0
   \end{bmatrix}.
\end{align}

\textbf{4. Operator \(\partial_r^2\) (second derivative):} A central difference scheme is adopted. The matrix \(D_2\) is defined as:
\begin{align}
     (D_2 u)_j \approx \frac{u_{j-1} - 2u_j + u_{j+1}}{\Delta r^2}, \quad j = 1, \dots, \bar{N},
\end{align}
   i.e.,
\begin{align}
     D_2 = \frac{1}{\Delta r^2}
   \begin{bmatrix}
   -2 & 1 & 0 & \cdots & 0 \\
   1 & -2 & 1 & \cdots & 0 \\
   0 & 1 & -2 & \ddots & \vdots \\
   \vdots & \ddots & \ddots & \ddots & 1 \\
   0 & \cdots & 0 & 1 & -2
   \end{bmatrix}.
\end{align}
Consequently, the Hamiltonian can be expressed as a matrix acting solely on coordinate space to evolve the wavefunction, while its entries depend on the color (singlet or octet) and the angular momentum quantum number \(l\) of the current state.
\subsubsection{The matrix form of the Lindblad operators}
Since the Lindblad operators involve transitions in color and angular momentum quantum numbers, we first need to define the color space and the angular momentum space. The color space, considering only the color singlet and color octet states, is two-dimensional. The angular momentum space is infinite-dimensional in theory, but owing to limited computational resources, it is often truncated to \(l_{\text{max}}\) in numerical simulations. The color singlet and color octet are given by
\begin{align}
     \ket{s} =  \begin{pmatrix} 1 \\ 0  \end{pmatrix} ,\quad \ket{o} = \begin{pmatrix} 0 \\ 1\end{pmatrix}.
\end{align}
Thus, the transition matrices between the color-singlet and color-octet states are given by
\begin{align}
    \ket{s}\bra{o} = \begin{bmatrix}
        0 & 1 \\ 0 & 0
    \end{bmatrix}, \qquad
    \ket{o}\bra{s} = \begin{bmatrix}
        0 & 0 \\ 1 & 0
    \end{bmatrix}.
\end{align}
The raising and lowering operators for angular momentum can be expressed as
\begin{align}
     \tilde{C}^{\uparrow} = \sum_{l=0}^{l_{\text{max}} - 1} \ket{l+1}\bra{l} ,\quad \tilde{C}^{\downarrow} = \sum_{l=1}^{l_{\text{max}}} \ket{l-1}\bra{l}.
\end{align}
Then, the six Lindblad operators are expressed as
\begin{align}
    L_1 &= \ket{o}\bra{s} \otimes   \sum_{l=0}^{l_{\text{max}} - 1} ( \ket{l+1}\bra{l} \otimes \gamma^{\uparrow}_{s\to o}   \overline{C}^{\uparrow}_{s\to o}),\label{L1}\\
    L_2 & = \ket{o}\bra{s} \otimes   \sum_{l=1}^{l_{\text{max}}} ( \ket{l-1}\bra{l} \otimes \gamma^{\downarrow}_{s\to o}   \overline{C}^{\downarrow}_{s\to o}),\\
     L_3 &= \ket{s}\bra{o} \otimes   \sum_{l=0}^{l_{\text{max}} - 1} ( \ket{l+1}\bra{l} \otimes \gamma^{\uparrow}_{o\to s}   \overline{C}^{\uparrow}_{o\to s}),\\
      L_4 &= \ket{s}\bra{o} \otimes   \sum_{l=1}^{l_{\text{max}}} ( \ket{l-1}\bra{l} \otimes \gamma^{\downarrow}_{o\to s}   \overline{C}^{\downarrow}_{o\to s}),\label{L4}\\
        L_5 &= \ket{o}\bra{o} \otimes   \sum_{l=0}^{l_{\text{max}} - 1} ( \ket{l+1}\bra{l} \otimes \gamma^{\uparrow}_{o\to o}   \overline{C}^{\uparrow}_{o\to o}),\\
         L_6 &= \ket{o}\bra{o} \otimes   \sum_{l=1}^{l_{\text{max}}} ( \ket{l-1}\bra{l} \otimes \gamma^{\downarrow}_{o\to o}   \overline{C}^{\downarrow}_{o\to o}),\label{L6}
\end{align}
 In the quantum circuit of this work, the qubit ordering strictly adheres to the convention that the total state is expressed as
\begin{align}
    \ket{\psi_{\text{total}}} = \ket{\text{aux}} \otimes \ket{\text{col}} \otimes \ket{l} \otimes \ket{r} = \ket{q_{n - 1} \dots q_2 q_1 q_0},
\end{align}
where the registers, from lowest to highest, are the coordinate-space register, the angular-momentum register, the color register, and the auxiliary register, and \(n\) is the total number of qubits in the circuit.

\section{Circuit Gate Complexity: Original vs. New Algorithm}\label{Circuit Gate Complexity: Original vs. New Algorithm}
To compare the gate complexity fairly, we compare the Trotterized implementations of the original and optimized algorithms. The original algorithm implements
\begin{align}
     &\prod_{i=1}^{m} e^{-iJ_i\sqrt{\delta t}},\notag \\
 J_i=|i\rangle\langle 0|&\otimes L_i+|0\rangle\langle i|\otimes L_i^\dagger,
\end{align}
on \(n_{\rm sys}+n_{\rm aux}\) qubits, where \(n_{\rm aux}=\lceil\log_2(m+1)\rceil\). The optimized algorithm implements
\begin{align}
    &\prod_{i=1}^{m} e^{-i\bar J_i\sqrt{\delta t}},\notag \\
    \bar J_i=|1\rangle\langle 0|&\otimes L_i+|0\rangle\langle 1|\otimes L_i^\dagger,
\end{align}
on \(n_{\rm sys}+1\) qubits, resetting the single ancilla after each factor.
For an arbitrary \(n\)-qubit Hamiltonian
\begin{align}
    H=\sum_{k=1}^{q}\beta_k P_k,
\end{align}
where each \(P_k\) is a tensor product of Pauli operators on \(n\) qubits. The number of Pauli terms \(q\) can be as large as \(O(4^n)\) in the worst case; for local or structured Hamiltonians, however, \(q\) may be polynomial.

Since different Pauli terms generally do not commute, the evolution operator \(e^{-iHt}\) must be implemented via Trotter decomposition. For the first-order Trotter formula,
\begin{align}
    e^{-iHt}
\approx
\left(
\prod_{k=1}^{q}e^{-i\beta_k P_k \delta t}
\right)^{s},
\qquad
\delta t=\frac{t}{s},
\end{align}
where \(s\) is the number of Trotter steps.

Each Trotter step requires sequentially implementing the exponentials \(e^{-i\beta_k P_k\delta t}\) for all \(q\) Pauli terms. A Pauli string of weight \(w_k\) can be exponentiated using \(O(w_k)\) CNOT gates via a standard CNOT ladder \cite{Sawaya:2020bky}. Since \(w_k\le n\), the number of CNOT gates per Trotter step is at most
\begin{align}
     O(qn).
\end{align}
Over \(s\) steps, the total number of CNOT gates is
\begin{align}
    O(s q n).
\end{align}
According to Eq.~(24) in Ref.~\cite{Childs:2019hts}, to simulate the first-order Trotter evolution with accuracy \(\epsilon\), it suffices to choose a number of steps \(s\) that is upper bounded by
\begin{align}
    s = O\!\left( \frac{ (\sum_k\|H_k\|t)^2}{\epsilon}\right),
\end{align}
where \(\sum_{k}\|H_k\| = \sum_{k=1}^{q}\|\beta_k P_k\| = \sum_{k=1}^{q}|\beta_k|\). Therefore, using a first-order Trotter decomposition, the number of CNOT gates required to simulate \(e^{-iHt}\) to accuracy \(\epsilon\) can be expressed as
\begin{align}
      O\left(
  q n \frac{ (\sum_{k=1}^{q}|\beta_k|t)^2}{\epsilon}
  \right).
\end{align}
For the original algorithm, the CNOT gate complexity for implementing the evolution of \(e^{-iJ\sqrt{\delta t}}\) via first-order Trotter decomposition is
\begin{align}
   O\left(
 \sum_{i=1}^{m} q_{J_i} (n_{\mathrm{sys}} + \lceil \log_2(m + 1) \rceil  ) \frac{ \left(\sum_{k=1}^{ q_{J_i}}|\beta_{J_i k}|\sqrt{\delta t}\right)^2}{\epsilon}
  \right).
\end{align}
Assuming that the CNOT gate complexity required for each \(e^{-iJ_i \sqrt{\delta t}}\) is comparable, this can be rewritten as
\begin{align}
   O\left(
 m q_{J_i} (n_{\mathrm{sys}} + \lceil \log_2(m + 1) \rceil  ) \frac{ \left(\sum_{k=1}^{ q_{J_i}}|\beta_{J_i k}|\sqrt{\delta t}\right)^2}{\epsilon} \label{o complexity}
  \right).
\end{align}
Analogously, for the new algorithm, the CNOT gate complexity required for \(e^{-iJ\sqrt{\delta t}}\) is
\begin{align}
O\left(
m q_{\bar{J}_i} (n_{\mathrm{sys}} + 1 ) \frac{ \left(\sum_{k=1}^{ q_{\bar{J}_i}}|\beta_{\bar{J}_i k}|\sqrt{\delta t}\right)^2}{\epsilon}
\right) .\label{n complexity}
\end{align}
A comparison of Eq.~\eqref{o complexity} and Eq.~\eqref{n complexity} reveals that, provided \( q_{\bar{J}_i} \leq q_{J_i}\), the optimized algorithm reduces both the required qubit resources and the circuit gate complexity.

Regardless of whether the number of Pauli terms in the Hamiltonian scales exponentially as \(O(4^n)\) or polynomially as \(O(\mathrm{poly}(n))\) with respect to the number of qubits \(n\), the optimized algorithm always requires at most as many qubits as the original algorithm. Specifically, in the present encoding, the ancilla register is reduced from \(\lceil\log_2(m+1)\rceil\) qubits to a single qubit, and the Pauli decomposition of \(\bar{J}_i\) does not contain more terms than that of \(J_i\); hence, \(q_{\bar{J}_i} \leq q_{J_i}\) in this construction. Although the new algorithm requires \(m\) individual reset operations, incurring a total reset cost of \(O(m)\), this overhead is negligible compared to the dominant CNOT gate complexity of the simulation and therefore does not compromise the advantage of the new algorithm.

\section{Resource-Optimized Algorithm for Lindblad Evolution: A Mathematical Derivation}
Before proceeding, we emphasize that the resource-optimized algorithm and the original algorithm are derived from the same mathematical framework, and their derivations and proofs follow the same procedure; this section is included for reference only. Note that resetting an ancillary qubit involves two operations: discarding its degrees of freedom (mathematically equivalent to taking the partial trace over it) and then re-preparing it in the \(|0\rangle\) state. In the resource-optimized algorithm,
\begin{align}
    \bar{J}_j =   |0\rangle\langle 1| \otimes L_j^\dagger + |1\rangle\langle0| \otimes L_j .
\end{align}
The initial state is
\begin{align}
    \rho_0 = |0\rangle\langle0| \otimes \rho.\label{initial state}
\end{align}
After evolution, tracing out the ancillary system yields
\begin{align}
    \mathcal{E}_j(\rho) = \operatorname{Tr}_{\text{aux}}\bigl( e^{-i  \bar{J}_j\tau} \rho_0 e^{i  \bar{J}_j\tau} \bigr).
\end{align}
The Taylor expansion of \(e^{-i \bar{J}_j\tau}\) yields
\begin{align}
    e^{-i  \bar{J}_j\tau} = I - i\tau \bar{J}_j - \frac{\tau^2}{2} \bar{J}_j^2 + i\frac{\tau^3}{6} \bar{J}_j^3 + \cdots .
\end{align}
Therefore,
\begin{align}
    &e^{-i\bar{J}_j\tau} \rho_0 e^{i\bar{J}_j\tau} \notag \\
    &= \bigl(I - i\tau \bar{J}_j - \frac{\tau^2}{2} \bar{J}_j^2 + \cdots\bigr) \rho_0 \bigl(I + i\tau \bar{J}_j - \frac{\tau^2}{2} \bar{J}_j^2 + \cdots\bigr).
\end{align}
We expand and collect the terms of order \(\tau^0\), \(\tau^1\), and \(\tau^2\). For the zeroth-order term \(I \rho_0 I = \rho_0\), taking the partial trace over the ancillary system yields \(\rho\). For the first-order term \(-i\tau \bar{J}_j \rho_0 + i\tau \rho_0 \bar{J}_j\), the first-order contribution vanishes because \(\operatorname{Tr}_{\text{aux}}(|1\rangle\langle0|)=\operatorname{Tr}_{\text{aux}}(|0\rangle\langle 1|)=0\). The second-order terms arise from three contributions:
\begin{align}
    -\frac{\tau^2}{2} \bar{J}_j^2 \rho_0 - \frac{\tau^2}{2} \rho_0 \bar{J}_j^2 + \tau^2 \bar{J}_j \rho_0 \bar{J}_j,\label{second-order terms}
\end{align}
where
\begin{align}
    \bar{J}_j^2 =  |0\rangle\langle0| \otimes L_j^\dagger L_j +  |1\rangle\langle 1| \otimes L_j L_j^\dagger.
\end{align}
Substituting the initial state given by Eq.~\eqref{initial state} into Eq.~\eqref{second-order terms} and taking the partial trace over the ancillary system yields
\begin{align}
    \tau^2  L_j \rho L_j^\dagger - \frac{\tau^2}{2}  \bigl(L_j^\dagger L_j \rho + \rho L_j^\dagger L_j\bigr).
\end{align}
After a single evolution under \(\bar{J}_j\), we finally obtain
\begin{align}
    \mathcal{E}_j(\rho) = \rho + \tau^2 L_j \rho L_j^\dagger - \frac{\tau^2}{2} \bigl(L_j^\dagger L_j \rho + \rho L_j^\dagger L_j\bigr) + O(\tau^4).
\end{align}
Finally, sequentially composing the single-step channels \(\mathcal E_j\) generated by the \(\bar J_j\), with the ancillary qubit reset after each step, yields
\begin{align}
    &\mathcal{E}(\rho) =  \mathcal E_m\circ\cdots\circ\mathcal E_1.\notag \\
    &= \rho + \tau^2 \sum_j L_j \rho L_j^\dagger - \frac{\tau^2}{2} \sum_j \bigl(L_j^\dagger L_j \rho + \rho L_j^\dagger L_j\bigr) + O(\tau^4).
\end{align}
This is precisely the dissipative part of the Lindblad equation Eq.~\eqref{Lindblad in reduce wavefunction} when evolved for a time \(\tau = \sqrt{\delta t}\). It also shows that the evolution generated solely by \(J\) (after tracing out the ancillary system) over a short time \(\tau\) yields a dissipative term proportional to \(\tau^2\). Therefore, to obtain the correct \(O(\delta t)\) dissipation in each small interval \(\delta t\), one must take \(\tau = \sqrt{\delta t}\).
\end{document}